\documentclass[prl,reprint,twocolumn,superscriptaddress,floatfix,noshowpacs,notitlepage,longbibliography]{revtex4-1}%
\usepackage{graphicx,bm,times}
\usepackage{amsmath}
\usepackage{amsfonts}
\usepackage{amssymb}
\usepackage{color}
\usepackage{hyperref}
\hypersetup{
	colorlinks = true,
	allcolors = {blue}
}

\begin{document}

\title{Formation of Hubbard-like bands as a fingerprint of strong electron-electron interactions in FeSe}

\author{Matthew D. Watson}
\email[corresponding author:]{matthew.watson@diamond.ac.uk}
\affiliation{Diamond Light Source, Harwell Campus, Didcot, OX11 0DE, UK}

\author{Steffen Backes}
\affiliation{Institut f\"{u}r Theoretische Physik, Goethe-Universit\"{a}t Frankfurt, Max-von-Laue-Str. 1, D-60438 Frankfurt am Main, Germany}
 
\author{Amir A. Haghighirad}
\affiliation{Clarendon Laboratory, Department of Physics,
University of Oxford, Parks Road, Oxford OX1 3PU, UK}

\author{Moritz Hoesch}
\affiliation{Diamond Light Source, Harwell Campus, Didcot, OX11 0DE, UK}

\author{Timur K. Kim}
\affiliation{Diamond Light Source, Harwell Campus, Didcot, OX11 0DE, UK}

\author{Amalia I. Coldea}
\affiliation{Clarendon Laboratory, Department of Physics,
	University of Oxford, Parks Road, Oxford OX1 3PU, UK}

\author{Roser Valent\'{i}}
\affiliation{Institut f\"{u}r Theoretische Physik, Goethe-Universit\"{a}t Frankfurt, Max-von-Laue-Str. 1, D-60438 Frankfurt am Main, Germany}

\begin{abstract}
We use angle-resolved photo-emission spectroscopy (ARPES) to explore the
electronic structure of single crystals of FeSe over a wide range of binding
energies and study the effects of strong electron-electron correlations. We
provide evidence for the existence of ``Hubbard-like bands" at high binding energies
consisting of
incoherent many-body excitations originating from Fe $3d$ states in addition to
the renormalized quasiparticle bands near the Fermi level. Many high energy
features of the observed ARPES data can be accounted for when incorporating effects
of strong local Coulomb interactions in calculations of the spectral function
via dynamical mean-field theory, including the formation of a 
Hubbard-like band. 
 This shows that over the energy scale of several eV,
local correlations arising from the on-site Coulomb repulsion and Hund's
coupling are essential for a proper understanding of the electronic structure
of FeSe and other related iron based superconductors. 

\end{abstract}
\date{\today}
\maketitle


{\it Introduction.-}
Understanding the role of electron-electron correlations in materials
exhibiting high-$T_c$ unconventional superconductivity is one of the central
problems within the field of strongly correlated electron systems. Unlike the
cuprates, the parent compounds of the Fe-based superconductors (e.g. LaFeAsO)
are not Mott insulators but antiferromagnetic metals at low temperatures, away
from half-filling. Nevertheless, local electron-electron interactions on the Fe
site do play an important role, although in this case it has been shown that
it is the Hund's coupling $J_H$ rather than the Coulomb repulsion $U$ which is
most important both for the magnetic ordering \cite{Johannes2009}  and
for the degree of band renormalization \cite{Haule2009,Aichhorn2009,Medici2011,Yin2011,Yu2011,
Ferber2012a,Ferber2012b,Georges2013,Backes2015,Fanfarillo2015}. From an
experimental point of view, clear manifestations of the effect of strong
correlations in Fe-based superconductors are found in enhancements of
quasiparticle effective masses deduced from specific heat
\cite{Hardy2016} and quantum oscillations measurements \cite{Putzke2012},
and from band renormalisations observed in Angle-Resolved Photo-Emission
Spectroscopy (ARPES) \cite{Borisenko2010,Evtushinsky2014,Maletz2014}. These
measurements indicate that the low-energy electronic structure broadly resembles
that predicted by Density Functional Theory (DFT) calculations, at least, at
temperatures above any magnetic or orbital orderings, but with the experimental
band dispersions being renormalised by a factor typically of $\sim$3
\cite{Borisenko2010,Evtushinsky2014}, although this varies substantially
between systems, and is orbital-dependent \cite{Yin2011}. However, while
general considerations of many-body theory would suggest that this band
renormalisation must be accompanied by the transfer of spectral weight into
incoherent excitations at higher binding energies \cite{Comin2013}, the high
energy spectral weight has only rarely been experimentally investigated in
Fe-based superconductors \cite{Evtushinsky2014,Nekrasov2015}. 

FeSe provides an ideal case to study the effect of strong correlations in
Fe-based superconductors. The recent availability of high-quality single
crystals \cite{Bohmer2014correct,Watson2015a} and thin films \cite{Tan2013} of FeSe 
has led to a surge of experimental work, including recent ARPES studies with a
focus on the origin of the nematic phase
\cite{Watson2015a,Zhang2015,Suzuki2015,Watson2015c,Watson2016a}. ARPES \cite{Watson2015a,Maletz2014},
quantum oscillations \cite{Watson2015a,Watson2015b,Terashima2014} and specific
heat measurements of FeSe \cite{Bohmer2014correct} have previously reported
significant orbital-dependent effective mass renormalisations. Theoretically, a significant
effect of correlations in FeSe has been found in combined Density Functional
Theory with Dynamical Mean Field Theory
(DFT+DMFT) calculations \cite{Yin2011,Aichhorn2010,Leonov2015}, in which local
Coulomb repulsion $U$ and Hund's coupling $J_H$ on the Fe site are accounted
for. 

In this paper, we present systematic ARPES studies of the spectral function of
FeSe to high binding energies. In addition to the renormalised quasiparticle
bands near the Fermi level, we find much broader features lying in a range of
1-2.5~eV binding energy, well separated from the quasiparticle structure and
the Se 4p bands at $\sim$3-6~eV. A ``peak-dip-hump" structure on such an
energy scale is usually a trademark of strong electron-electron interactions,
which reduce the spectral weight of the quasiparticle peak and give rise to Hubbard
bands at higher and lower binding energies~\cite{Zhang1993}.
Our DFT+DMFT calculations are able to reproduce many of the qualitative
features of the experimental electronic structure at high binding energies, including the formation of
Hubbard-like bands of incoherent spectral weight. While accounting for local
electron-electron interactions within DFT+DMFT alone is not sufficient for
a perfect description of the experimental Fermi surface, we show that the
strong interactions are responsible for the overall form of the spectral
function of FeSe over an energy scale of several eV. 

{\it Methods.-}
Single crystals of FeSe were grown by the vapor-transport method
\cite{Watson2015a}. ARPES measurements were performed at the I05 beamline at
Diamond Light Source at temperatures below 10~K. 
%
%
ARPES measurements are a probe of the one-particle spectral function $A(\omega,\mathbf{k})$ \cite{Comin2013}, multiplied by the Fermi occupation function and the matrix elements for photo-emission \cite{Comin2013}, with some additional background. This spectral function is commonly expressed as:

\begin{equation}
A(\omega,\mathbf{k}) = - \frac{1}{\pi}\frac{\Sigma''(\omega{},\mathbf{k})}{[\omega+\mu-\epsilon^b_{\mathbf{k}}-\Sigma'(\omega{},\mathbf{k})]^2 + [\Sigma''(\omega{},\mathbf{k})]^2}
\end{equation}

where $\epsilon^b_{\mathbf{k}}$ is the bare non-interacting dispersion, $\mu$ the chemical potential and $\Sigma'$ and $\Sigma''$ are the real and imaginary parts of the self-energy, which in general is orbital-, frequency- and momentum-dependent. In many materials where electronic correlations are weak and do not play a significant role, $\Sigma$ is small and sharp dispersions can be observed in ARPES measurements to binding energies of several eV, usually in good agreement with the DFT dispersions. On the other hand, in FeSe, electron-electron interactions on the Fe $3d$ site do give a significant contribution to the self-energy~\cite{Aichhorn2010,Leonov2015}, while the system remains metallic. Therefore, the observed dispersions close to the Fermi level at low temperatures can be interpreted as coherent quasiparticles with renormalised dispersions $\epsilon^q_{\mathbf{k}}=\epsilon^b_{\mathbf{k}}+\Sigma'$, and a scattering rate  $\Sigma''$ that introduces a finite lifetime for quasiparticle excitations. Depending on the form of  $\Sigma(\omega,\mathbf{k})$ there may be apparent ``kinks" or ``waterfalls" \cite{Iwasawa2012} in the spectral function where the observed states transform from the renormalized quasiparticle peak close to the Fermi level into incoherent excitations at higher
or lower binding energies. Generally speaking, at higher binding energy,
features can become very broad and incoherent when $\Sigma''$ becomes large,
and in particular the formation of Hubbard-like bands is possible
\cite{Kotliar2004,Kotliar2006}. 
While experimental evidence of Hubbard bands has been largely reported  
for effective one-band systems~\cite{Golden2001,Comin2013}, results for multiorbital
systems are scarce with only a few well-studied exceptions like
transition metal oxides~\cite{Inoue1995,Takizawa2009,Aizaki2012,Backes2016}.

The DFT+DMFT calculations were
performed within the local density approximation in DFT and using the full-potential linear augmented plane-wave (FLAPW) basis
within the WIEN2k~\cite{Wien2k} package. 
Calculations were done for the 
orthorhombic crystal structure~\cite{Khasanov2010}, and 
differences in the calculation to the tetragonal crystal structures
were small (Supplemental Material, SM \footnote{See Supplemental Material at [URL will be inserted by publisher] for further experimental data and details of DMFT calculations performed with different interaction parameters.}).               
We used the projection method onto a local basis as described in Refs.~\cite{Aichhorn2009,Ferber2014}, with a window 
 encompassing both the iron $3d$ and selenium $4p$ states. 
 The impurity
problem for the Fe $3d$ orbitals was solved with the strong-coupling
continuous-time quantum Monte-Carlo method~\cite{Gull2011} using the ALPS
package~\cite{Bauer2011}. As interaction parameters we use the established
values of $U$=4 eV, $J_H$=0.8~eV~\cite{Aichhorn2010,Miyake2010}. We employed
the fully-localized limit~\cite{Anisimov1993,Dudarev1998} for the
double counting term, and the stochastic analytic continuation method for obtaining
real-frequency data~\cite{Beach2004}. Calculations were performed at a
temperature of $\beta$=100 eV$^{-1}$, corresponding to $T=$116 K.

\begin{figure}
	\centering
	\includegraphics[width=\linewidth]{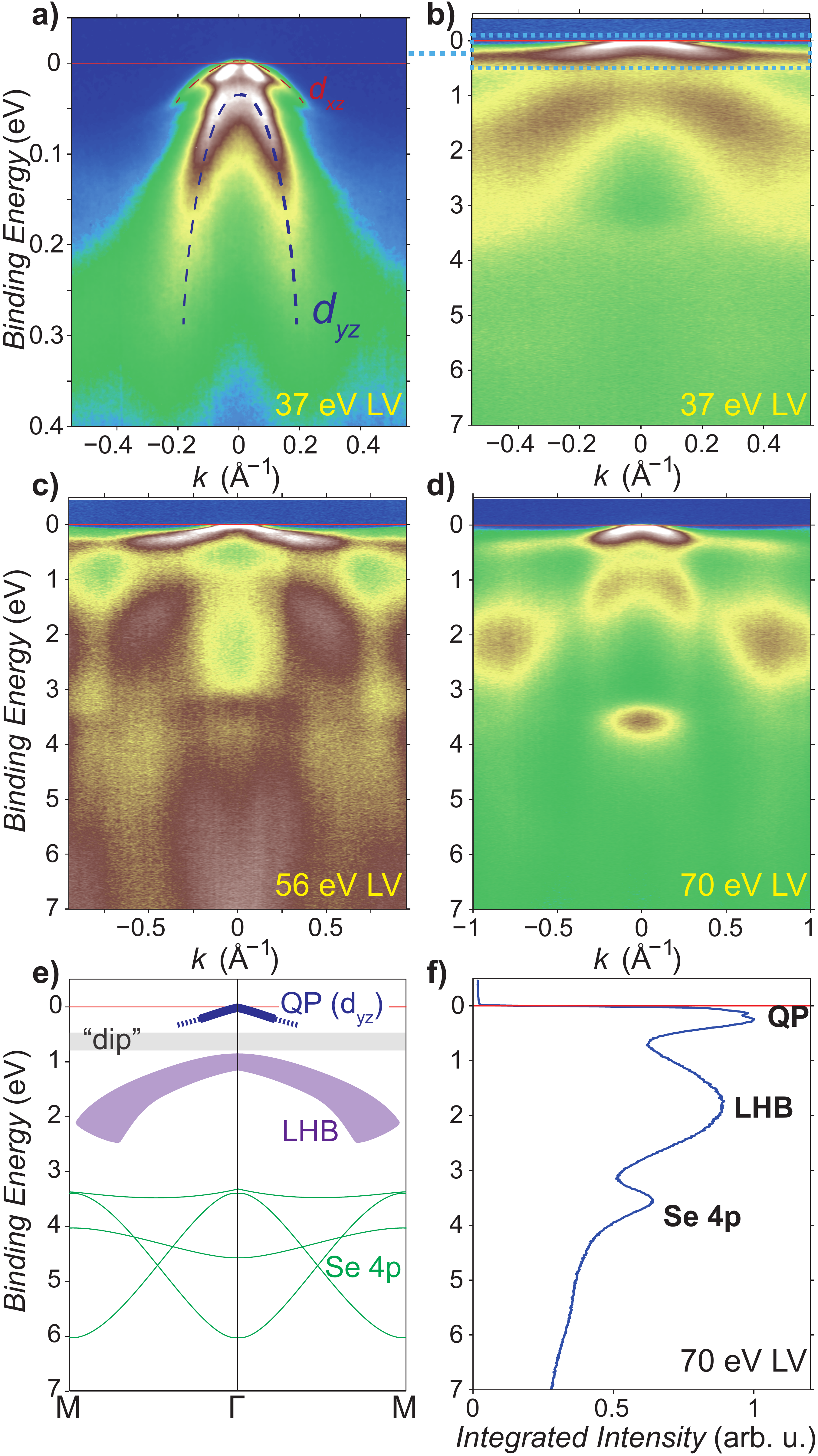}
	\caption{a-d) ARPES data in the M-$\Gamma$-M direction at 37 eV in linear vertical (LV) polarisation at 10~K. In this geometry a hole-like quasiparticle band with $d_{yz}$ orbital character dominates the photoemission spectrum. b-d) Measurements in the same geometry at different incident photon energies. The data extend to high binding energies, where much broader features are found. e) Schematic of the high energy spectrum. f) Integrated spectral weight from panel (d), showing features associated with the quasiparticle (QP), lower Hubbard band (LHB) intensities as well as a contribution from the Se $4p$ bands.}
	\label{fig1}
\end{figure}

\begin{figure*}
	\centering
	\includegraphics[width=\linewidth]{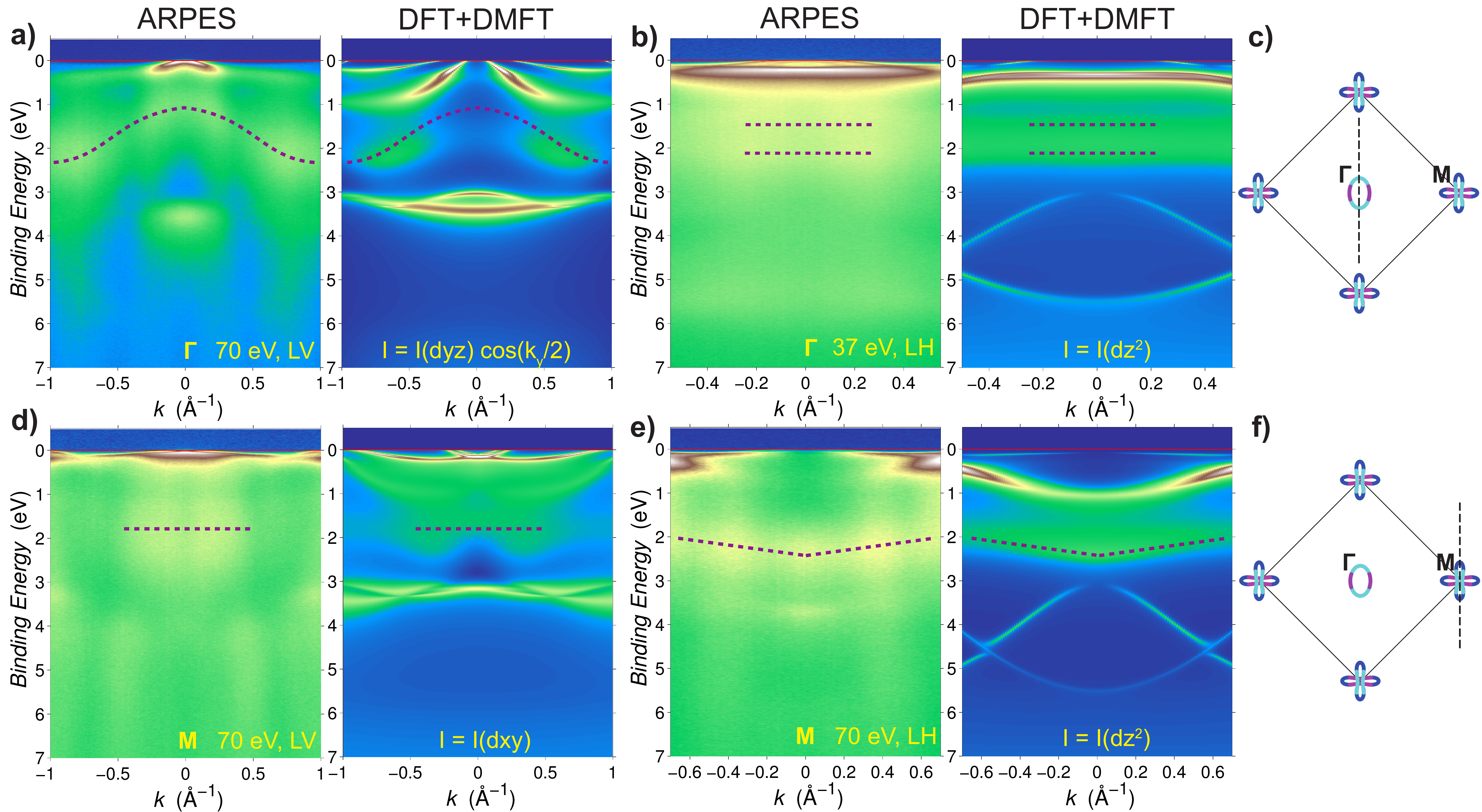}
	\caption{a,b,d,e) Comparison of ARPES spectra with DFT+DMFT calculations.
  The DFT+DMFT simulations are obtained by applying simple
selection rules to the orbitally-resolved spectral weight, to account for the experimental matrix elements effects. Dashed lines are guides to the eye showing the location of experimental incoherent Fe $3d$ spectral weight. c,f)
Schematic measurement geometries of the cuts shown in panels (a-b) (and also
Fig.~\ref{fig1}) and (d,e) respectively.} 
	\label{fig2}
\end{figure*}

{\it Results.-}
In Fig.~\ref{fig1} we present high-symmetry ARPES measurements for FeSe in the
M-$\Gamma$-M direction, using linear vertical (LV) polarisation. In this
geometry, strong matrix elements effects dictate that the spectral weight
arises overwhelmingly from a single hole-like band with $d_{yz}$ character \cite{Watson2015a},
which simplifies the observation. Fig.~\ref{fig1}a) focuses on the dispersion
of this $d_{yz}$ hole band close to $E_F$. The quasiparticle band dispersions
undergo $\sim$20 meV band shifts in the nematic phase  \cite{Watson2016a}, but
these are very small perturbations on the energy scales of a few eV as considered in this
paper. Due to spin-orbit coupling there is a small mixing of spectral
weight onto the outer ($d_{xz}$) hole band near the Fermi level
\cite{Watson2015a}. In Figs.~\ref{fig1}b), c), d) we present measurements extending to
 binding energies of 7 eV at a selection of incident photon energies. Varying the
photon energy has multiple effects. Firstly,  the $k_z$ of the slice of the
Brillouin zone  probed  varies (e.g. 37 eV and 56 eV are near $\Gamma$ and Z
points respectively \cite{Watson2015a}) which can affect the position and
orbital character of bands. Secondly ARPES matrix elements themselves have a
complex photon-energy dependence. Finally, if the photon energy passes through
an Fe or Se resonance this may affect the relative intensity of Fe or Se
contributions to the photoemission (this gives an enhancement of the Se bands
in the 56 eV spectra  in Fig.~\ref{fig1}c)). We do not attempt to disentangle
all these effects which lead to the differences between spectra presented
in  Figs.~\ref{fig1} b-d), but
rather point out five common features which are observed at all photon
energies, as we have represented schematically in Fig.~\ref{fig1}e,f); (i)
near the Fermi level the observed band is both shifted and renormalised with
respect to DFT calculations, as has been widely reported in Fe-based
superconductors \cite{Carrington2011,Borisenko2015}, although the
much smaller than expected Fermi surfaces in FeSe is a unique feature.
(ii) The quasiparticle band dispersions become much sharper towards the
Fermi level. (iii) There is generally a dip in intensity in the range
$\sim$0.5-1 eV in experiments, where neither quasiparticles nor incoherent
excitations are found. (iv) traces of the Se $4p$ bands are detected in
the range 3-6 eV binding energy, as predicted by DFT. Therefore, the Se $4p$
bands do not experience any significant renormalisation.  Finally, (v) in the range
of $\sim$1-2.5 eV we observe an anomalous broad band of intensity which 
cannot 
be attributed to either a Fe-$3d$ quasiparticle band or a Se $4p$ band.
The width of this spectral feature is of the order of $\sim$1~eV which indicates that these
excitations are very short lived. 
We interpret this as a ``Hubbard-like band", consisting of incoherent spectral weight that is a precursor of the localized electron-removal states, the lower Hubbard band, in Mott-Hubbard-insulating systems. 
No significant temperature-dependence was found in the high energy features up
to 150~K (SM). 

In Fig.~\ref{fig2} we present a selection of ARPES spectra obtained in
different measurement conditions, which indicate that this incoherent spectral
weight in the region around $\sim$1-2.5~eV is a general feature of FeSe,
 and not specific to a particular band or
geometry.
 Next to each
experimental measurement, we also show how the high energy features of FeSe
seen by ARPES can be qualitatively reproduced by calculations of the spectral
function in DFT+DMFT. In order to perform a comparison to ARPES data, simple
selection rules are employed to simulate the photoemission matrix elements in
that geometry. 
They
are based on both symmetry
considerations and the identified orbital character of the primary
quasiparticle bands in the cut \cite{Watson2016a} \footnote{We note that
experimental measurements have an additional background which will not be
accounted for in the calculation, which also distorts the colors scales}.  As presented in Fig.~\ref{fig2}a)
  DFT+DMFT reproduces the observed renormalised quasiparticle
$d_{yz}$ band and some additional high energy spectral weight around 1-2.5 eV.
However, the agreement is not perfect, and the renormalisation of the effective masses in DFT+DMFT (e.g. $m^*/m_{LDA} = $~2.09 for $d_{xz/yz}$, SM) is less than the experiments ($\sim$2-4 for $d_{xz/yz}$ bands, \cite{Watson2015a}), which is to be expected due to the
neglect of spin-flip and pair-hopping terms~\cite{Aichhorn2010} and dynamical
screening effects~\cite{Casula2012,Biermann2014}. Still, we expect that any Hubbard
band-like features are not qualitatively affected by these approximations, since
their binding energy is governed by the low-energy static values of the interaction,
which are accounted for in the calculation.
In Fig.~\ref{fig2}b) the DFT+DMFT
calculation shows a broad band of incoherent $d_{z^2}$ spectral weight in good
correspondence with anomalous weight found in ARPES around 1.5-2.5 eV. In Fig.~\ref{fig2}d), DFT+DMFT finds some incoherent spectral weight in the
$d_{xy}$ orbital around the M point, similar to the ARPES data. Finally in Fig.~\ref{fig2}e) the $d_{z^2}$ weight through the M point is reproduced very well, showing a clear formation of a Hubbard-like band. Overall there is a qualitative
good agreement between calculations and experiment, as the DFT+DMFT
 technique correctly captures both the
renormalised quasiparticle bands which sharpen approaching the Fermi level,
along with the incoherent spectral weight around 1-2.5 eV.

\begin{figure}[t]
	\centering
	\includegraphics[width=\linewidth]{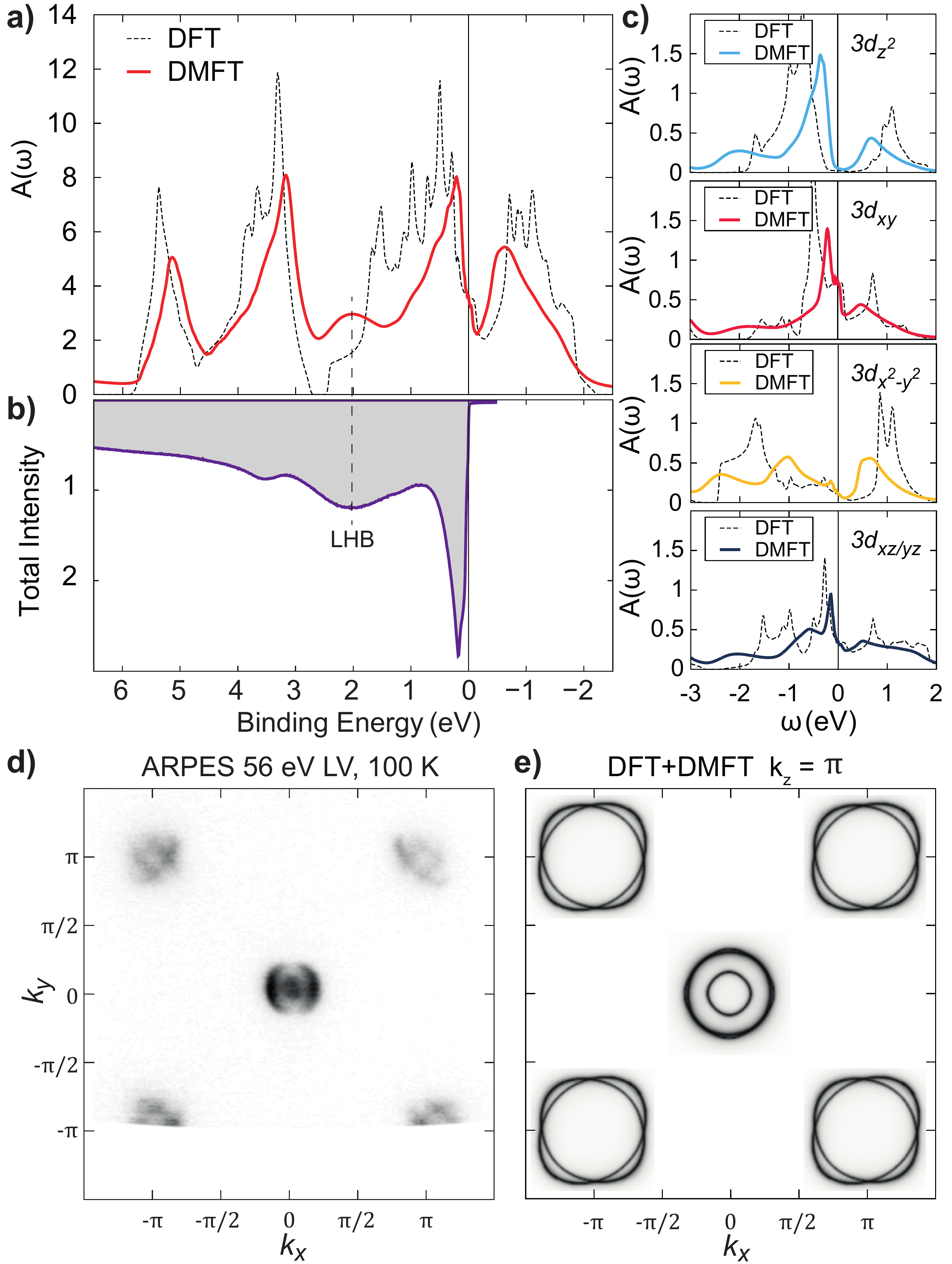}
	\caption{a) Integrated spectral weights as calculated from DFT
and DFT+DMFT. b) Total intensity obtained by summation over the different cuts in Fig.~\ref{fig2}a,c,d). c) Orbital-resolved spectral weight. d) Comparison of Fermi surfaces determined experimentally at 100~K in the tetragonal phase and e) as calculated by DFT+DMFT projected in the $k_z=\pi$ plane. }
	\label{fig3}
\end{figure}

In Fig.~\ref{fig3}a) we compare the integrated spectral weight from our
DFT+DMFT calculation with the result from DFT. Notably the Fe $3d$ bandwidth
develops a peak-dip-hump structure which is not present in the DFT; this
arises from the separation of the quasi-particle bands and the Hubbard
satellite peak around 2 eV. As expected, the DMFT treatment does not strongly
affect the Se $4p$ bands. 
In Fig.~\ref{fig3}c) we show the different orbital
contributions to the total spectral weight. The Hubbard band feature appears
most clearly in the $d_{z^2}$ and $d_{xz/yz}$ orbitals but can be identified in
all, similar to Ref.~\cite{Aichhorn2010}. In Fig.~\ref{fig3}b) we compare the
total calculated spectral weight with a summation of the experimental data from
Fig.~\ref{fig2}a,c,d). Similar qualitative features are found, with good agreement on the position of the Hubbard-like peak, which
supports the chosen values of the interaction parameters $U,J_H$, which are also
close to values recently determined from first-principles calculations
\cite{vanRoekeghem2016}.  We note that in DFT+DMFT the Hubbard-like peak
shifts to higher binding energies with increasing $U,J_H$, where the Hund's coupling
$J_H$ has a stronger effect on the energy of the Hubbard band than
$U$ (SM).

Finally in Fig.~\ref{fig3}d,e) we compare the experimental Fermi surfaces of FeSe at
100 K with the calculated ones.  The measured Fermi surfaces are significantly shrunk compared to the prediction of DFT+DMFT. In order to match
the experimental dispersions, the real parts of the self-energies would 
need to be significantly momentum-dependent in order to introduce a
downward-shift for hole bands at the Gamma point and an upward-shift for the
electron bands at the M point~\cite{Borisenko2015,Watson2015a}. In DMFT, the
considered interactions ($U,J_H$) are purely local and the self-energies
$\Sigma(\omega)$ are independent of $\mathbf{k}$, albeit orbital-dependent, so
that momentum-dependent shifts of the DFT bandstructure can only result from
the momentum-dependent orbital characters of the bands. The limitations of
DFT+DMFT at the Fermi level indicate that effects not included in
the calculations such as non-local
inter-site interactions \cite{Jiang2016}, coupling to bosonic modes
\cite{Ortenzi2009} or frustrated magnetism \cite{Glasbrenner2015}
are likely to be relevant to the low-energy physics.
However, for the wide energy scales considered in this paper our DFT+DMFT
calculation is able to satisfactorily capture many of the high energy features of our ARPES
spectra, including the presence of incoherent spectral weight in the form of
Hubbard-like bands at high binding energies, with specific orbital-dependent
agreements. 
Our experiments and calculations place bulk FeSe as a
significantly correlated metal, with coherent quasiparticles at the Fermi
level, but also exhibiting incoherent spectral weight at high binding energies,
consistent with earlier photoemission studies \cite{Yamasaki2010}. 

{\it Conclusion.-}
%
To summarise, we have provided systematic experimental evidence, backed up by theoretical DFT+DMFT calculations, for the emergence of a Fe $3d$ Hubbard-like band in the spectral function of FeSe, distinct from the quasiparticle states near the Fermi level. This high-energy feature is interpreted as a fingerprint of the effect of strong electron-electron correlations.   
Despite the strong renormalisation and shift of spectral weight into the Hubbard-like features, a well-defined quasiparticle peak at the Fermi level is retained. 
Therefore FeSe provides a rare opportunity to study Hubbard-band physics in a significantly correlated, metallic, multiorbital system. 
The unique properties of FeSe continue to provide theoretical challenges, but
we have demonstrated that the DFT+DMFT technique captures the essential
features of the high-energy spectral function well, highlighting the importance
of local Coulomb interactions and Hund's coupling for both low and high energy features in Fe-based superconductors.

\begin{acknowledgments}
\section{Acknowledgments}
We thank S.~Biermann, B.~B\"uchner, D.~Guterding, H.~Iwasawa, H.~O.~Jeschke and L.~C.~Rhodes for useful discussions. 
 We acknowledge Diamond Light Source for time on beamline I05 under proposals CM12153, SI10203. S.B. and R.V thank the Deutsche Forschungsgemeinschaft (DFG)
for financial support through grant SPP 1458. Part of the work was supported by the EPSRC (EP/L001772/1,EP/I004475/1,EP/I017836/1). A.A.H. acknowledges the financial support of the Oxford Quantum Materials Platform Grant (EP/M020517/1). A.I.C. acknowledges an EPSRC Career Acceleration Fellowship (Grant No. EP/I004475/1).
\end{acknowledgments}
%
%

\clearpage

\end{document}